\documentclass[aps,superscriptaddress,twocolumn,showpacs,amsmath,amssymb]{revtex4}
\usepackage{graphicx,amsmath}
\input{epsf}

\begin{document}

\title{Fractal Weyl law for Linux Kernel Architecture} 

\author{L.Ermann}
\affiliation{\mbox{Laboratoire de Physique Th\'eorique du CNRS (IRSAMC), 
 Universit\'e de Toulouse, UPS, F-31062 Toulouse, France}}
\author{A.D. Chepelianskii}
\affiliation{LPS, Universit\'e Paris-Sud, CNRS, UMR8502, F-91405 Orsay, France}
\author{D.L.Shepelyansky}
\affiliation{\mbox{Laboratoire de Physique Th\'eorique du CNRS (IRSAMC), 
Universit\'e de Toulouse, UPS, F-31062 Toulouse, France}}

\date{May 9, 2010; Revised: September 15, 2010}

\begin{abstract}
We study the properties of spectrum and eigenstates 
of the Google matrix of a directed network
formed by the procedure calls in the Linux Kernel.
Our results obtained for various versions of the Linux Kernel
show that the spectrum is characterized by the fractal Weyl law
established recently for systems of quantum chaotic scattering and
the Perron-Frobenius operators of dynamical maps.
The fractal Weyl exponent is found to be
$\nu \approx 0.65$ that corresponds to the fractal dimension
of the network $d \approx 1.3$. An independent computation 
of the fractal dimension by the cluster growing method, generalized for directed networks, 
gives a close value $d \approx 1.4$.
The eigenmodes of 
the Google matrix of Linux Kernel 
are localized on certain principal nodes. We argue that the fractal Weyl law 
should be generic for directed networks with the fractal dimension $d<2$.
\end{abstract}

\pacs{89.75.Hc, 05.45.Mt, 07.05.Bx, 05.45.Df}

\maketitle

\section{Introduction}

The celebrated Weyl law \cite{weyl} determines the number of quantum states  
in a given energy interval of conservative Hamiltonian 
systems in a phase space dimension $d$ as a function 
of the Planck constant $\hbar$
\begin{equation}
N_E \propto \hbar^{-\nu}  \;, \; \nu = d/2 \; .
\label{eq1} 
\end{equation}
Recently this law has been generalized to the fractal Weyl law
applicable to 
open quantum systems with complex energy spectrum 
of nonunitary quantum operators 
\cite{sjostrand,zworski2003,schomerus,keating,dls2008,saraceno2009}.
In this case the exponent $\nu$ is linked to the
fractal dimension of invariant sets
of classical non-escaping orbits.
Such type of operators appear in physical
problems of quantum chaotic scattering,
open quantum maps, metastable quantum states in molecules and nuclei.

In addition to open quantum systems it has been shown that the
fractal Weyl law is also valid for the Perron-Frobenius
operators of classical dynamical systems \cite{mbrin}
with absorption or dissipation \cite{ermann}. 
A finite size matrix approximant ${\bf S}$
of such operators is efficiently constructed using
the Ulam method \cite{ulam} which divides the whole 
phase space on $N$ cells of size $\epsilon$.
The eigenvalues $\lambda_i$ of the Ulam matrix approximant ${\bf S}$
of the Perron-Frobenius operator are located in a complex plane of 
a circle with $|\lambda| \leq 1$.
For two-dimensional (2D) maps we have 
the matrix size $N \propto 1/\epsilon^2$
so that a cell area $\epsilon^2$ effectively plays a role of $\hbar$
and $N_\lambda \propto N^\nu$ where $N_\lambda$
is a number of eigenvalues $\lambda_i$ with
$|\lambda| \leq |\lambda_i| \leq 1$. For dynamical strange attractors
and 2D maps with absorption we have $\nu=d/2$
where $d$ is a fractal dimension of a dynamical invariant set,
e.g. a fractal dimension of a strange attractor \cite{ermann}. 
For $\nu <1$ the majority of eigenvalues have 
$|\lambda| \rightarrow 0$ for $N \rightarrow \infty$. 

In fact the Ulam method applied to 1D and 2D maps naturally generates
the Ulam networks \cite{zhirov} which properties can be rather similar to
the properties of scale-free directed networks such as the World Wide Web (WWW).
Such type of networks naturally appear in various fields of science
(see e.g. 
\cite{watts,newman,barabasi,dorogovtsev,googlebook,sornette,donatob}).
Therefore we can expect that the fractal Weyl law can appear also for 
directed networks similar to those of the WWW. The properties of such networks
are well characterized by the Google matrix ${\bf G}$ \cite{brin}
constructed on the basis of directed links between nodes:
\begin{equation}
  G_{ij} = \alpha S_{ij} + (1-\alpha) / N 
\label{eq2} 
\end{equation} 
where the matrix ${\bf S}$ is obtained by normalizing to 
unity all columns of the adjacency matrix,
and replacing columns with zero elements by $1/N$, 
$N$ being the network size \cite{googlebook}.
In the WWW context, the damping parameter $\alpha$
describes the probability 
to jump to any node for a random surfer. For $0< \alpha < 1$  
the only eigenvector with $\lambda=1$ is the PageRank
vector which has non-negative components and plays an important role
in ranking of the WWW sites \cite{brin,googlebook,donatob}.
The matrix ${\bf G}$ belongs to the class 
of Perron-Frobenius operators \cite{googlebook}.
The spectrum of ${\bf G}$
for the WWW university networks
was studied recently in \cite{georgeot}
and it was shown that a significant fraction of eigenvalues 
is concentrated at $|\lambda| \rightarrow 0$ that is
a characteristic property of spectra with the Weyl law.
However, no direct evidence for the fractal Weyl law
for the WWW was found there for the reasons we
explain below.
\begin{figure}[ht]
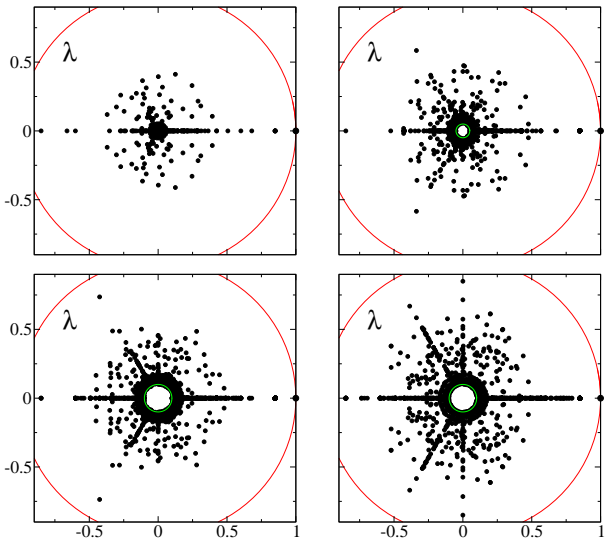

\begin{center}  
\includegraphics[width=.22\textwidth]{fig1a.eps}
\includegraphics[width=.22\textwidth]{fig1b.eps}
\includegraphics[width=.22\textwidth]{fig1c.eps}
\includegraphics[width=.22\textwidth]{fig1d.eps}
\caption{(Color online) 
Distribution of eigenvalues $\lambda$ in the complex plane 
for the Google matrix ${\bf G}$ of the 
Linux Kernel versions $2.0.40$ (top left panel),  
$2.4.37.6$ (top right panel) and $2.6.32$ 
(left bottom panel) with $N=14079$, $N=85756$ and $N=285509$ respectively. 
Bottom right panel shows eigenvalue distribution for 
the Google matrix ${\bf G^*}$ with inverted link directions in the PCN of 
the Linux Kernel version  $2.6.32$ with $N=285509$.
Solid lines represent the unit circle and 
the lowest limit of computed eigenvalues in top right and bottom panels. 
}
\label{fig1}
\end{center}
\end{figure}

It is interesting to note that many complex networks 
are self-similar and have fractal properties (see e.g.
\cite{fractal,havlinnphys} and Refs. therein).
Thus it was shown that the fractal dimension of 
networks of WWW and actors is $d= 4.1, 6.3$ respectively \cite{fractal}.
It is argued that the origin of fractal
architecture of network is  linked to a strong effective 
repulsion between the most connected nodes \cite{havlinnphys}.
Such a generic fractal structure of complex networks
also indicates that under certain conditions
the fractal Weyl law can appear in
the spectrum of the Google matrix of such networks.
However, it is also important to note that
usually the fractal properties are investigated
for undirected networks or by converting directed
networks into undirected ones (see e.g.  \cite{fractal,havlinnphys}).
Here, on a concrete example of the Linux Kernel network
we show that the directionality of the network
places a crucial role so that the directed network
has different fractal dimension compared to
the converted undirected network with the same nodes.
The importance of directionality is very clear from the
spectrum of the Google matrix: the spectrum is in a complex plane 
for a directed network while for an undirected network the spectrum
is on a real line. 

The paper is composed as following:
the specrum of the Google matrix is analyzed in Section II,
the fractal properties of the Linux Kernel network are considered
in Section III, the properties of Google matrix eigenstates
are studied in Section IV and discussion of the results is 
given in Section V.

\section{Spectrum of Google matrix}

To check the validity of the fractal Weyl law for a certain type of scale-free
networks we focus on the Procedure Call Network (PCN)
of the Linux Kernel software  introduced and studied recently in
\cite{alik}. In the PCN the directed links are formed by the calls between
procedures and the Google matrix (\ref{eq2})
is constructed in the same way as for the WWW. The results of \cite{alik}
show that only about $1\%$ of eigenvalues have $|\lambda| > 0.1$
that is by a factor 50 smaller compared to the university networks
of similar size
studied in \cite{georgeot}. This gives a significant indication on
appearance of the fractal Weyl law in the PCN. To study the
spectral properties of ${\bf G}$
we use the data for the Google matrix at $\alpha=0.85$
obtained in \cite{alik}. Our results are not sensitive to $\alpha$
which affects the spectrum of ${\bf G}$ 
in a rather simple way \cite{googlebook,zhirov}.
\begin{figure}[t]
\begin{center}  
\includegraphics[width=.475\textwidth]{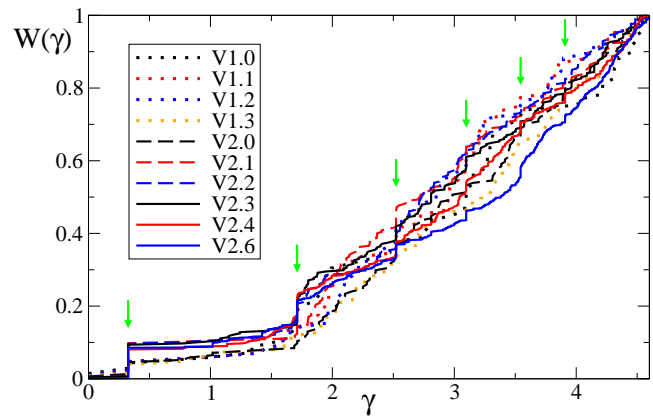}
\caption{(Color online) 
Dependence of the integrated density of states $W$
on the relaxation rate
$\gamma=-2\ln{(\lambda)}$ for the Google matrix 
of different versions of Linux Kernels 
from $V1.0$ with $N=14079$ to $2.6.32$ 
with $N=285509$ as it is shown on the legend.
The highly degenerate values of $\gamma$ are 
marked by arrows at
$\gamma_m=-2\ln{(\alpha/m)}$ for $m=1,2,3,4,5,6$.
}
\label{fig2}
\end{center}
\end{figure}

\begin{figure}[h!]
\begin{center}  
\includegraphics[width=.475\textwidth]{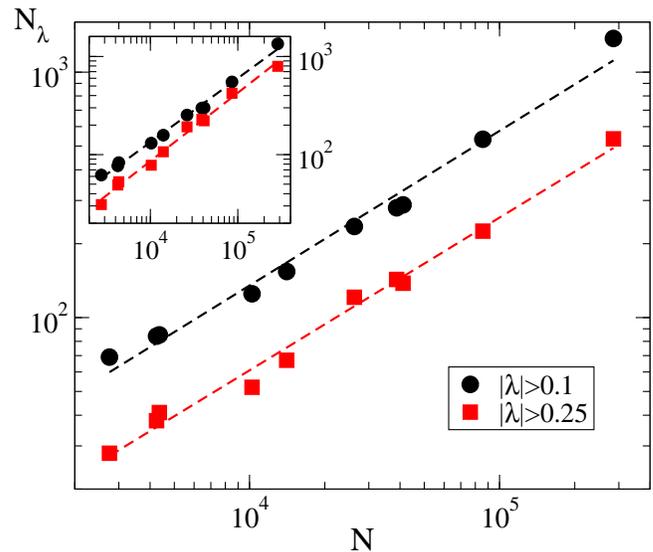}
\caption{(Color online) Dependence of the integrated number 
of eigenvalues $N_\lambda$ with 
$\vert\lambda\vert>0.25$ (red/gray squares) 
and $\vert\lambda\vert>0.1$ (black circles)
as a function of the total number of processes 
$N$ for versions of Linux Kernels. 
The values of $N$ correspond (in increasing order) to Linux Kernel versions 
$1.0$, $1.1$, $1.2$, $1.3$, $2.0$, $2.1$, $2.2$, $2.3$, $2.4$ and $2.6$.
The power law $N_\lambda\propto N^{\nu}$ has fitted values 
$\nu_{\vert\lambda\vert>0.25}=0.622 \pm 0.010$ and
$\nu_{\vert\lambda\vert>0.1}=0.630 \pm 0.015$.
Inset shows data for the Google matrix ${\bf G^*}$ 
with inverse link directions,
the corresponding  exponents are 
$\nu^*_{\vert\lambda\vert>0.25}=0.696 \pm 0.010$ and
$\nu^*_{\vert\lambda\vert>0.1}=0.652 \pm 0.007$. 
}
\label{fig3}
\end{center}
\end{figure}
The eigenvalues $\lambda_i$ and the right eigenvectors $\psi_i(j)$
satisfy the equation $\sum_{j'} G_{jj'} \psi_i(j') = \lambda_i \psi_i(j)$.
To obtain the spectrum of the PCN with large sizes $N$
we used the Arnoldi method from ARPACK
library \cite{arnoldi} that allowed to find all
eigenvalues with $|\lambda| >0.1$ up to the maximum
$N=285509$. The complex spectrum of $\lambda_i$
is shown in Fig.~\ref{fig1} for three versions of Linux Kernel.
With the increase of the version number and the size $N$
certain characteristic structures appear in the distributions of 
$\lambda_i =|\lambda_i|\exp(i\varphi_i)$
in the complex plane.  Thus, there are clearly visible 
lines at real axis and 
polar angles $\varphi = \pi/2, 2\pi/3, 4\pi/3, 3\pi/2$.
The later are related to certain cycles in procedure calls,
e.g. an eigenstate at $\lambda_i =0.85 \exp(i 2\pi/3)$ is 
located only on 6 nodes.
In Fig.~\ref{fig1} we also present
the spectrum of  the Google matrix ${\bf G^*}$
which is obtained by an inversion of link directions
in the adjacency matrix of the PCN as it was proposed in \cite{alik}.
It is characterized by a similar spectrum structure 
but a  larger radial distribution of $\lambda_i$.

The dependence of the normalized integrated density of states $W(\gamma)$,
with eigenvalues $0 \leq \gamma_i \leq \gamma$,
on the eigenvector relaxation rate $ \gamma$
is shown in Fig.~\ref{fig2}. Here $\gamma_i=-2 \ln |\lambda_i|$ and 
$W(\gamma)$ is normalized on the number of states 
$N_\lambda$ in the interval
$\lambda \leq |\lambda_i| \leq 1$ with $\lambda=0.1$.
In average the dependence $W(\gamma)$ is the same for all 10 versions
of Linux Kernel 
even if there are certain fluctuations.
There are clear vertical steps in $W(\gamma)$ at 
$\gamma_m=-2\ln(\alpha/m)$ which are due to a presence of many degenerate
states at the corresponding values of $\lambda$ (e.g. 116
for $\lambda=\alpha$ for the version 2.6.32). Such degeneracy is known to be
typical of the WWW networks (see e.g. \cite{donatob,georgeot}).
\begin{figure}[h!]
\begin{center}  
\includegraphics[width=.475\textwidth]{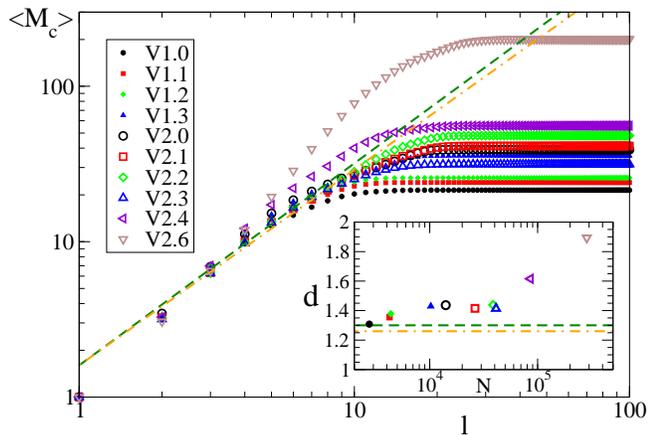}
\caption{(Color online) Cluster growing fractal dimension 
of the Linux Kernel directed networks. 
The average mass $\langle M_c\rangle$, with uniformly 
distributed seed, is shown for 
different Kernel versions as a function 
of the \emph{network distance} $l$, in logarithmic scale.
The corresponding fractal dimensions $d$ for all versions are 
shown in the inset panel, where the power law fit is done for 
the range $1 \leq l \leq 10$.
The dashed and dot--dashed lines (in both panels) represent 
the power law dependence 
$\langle M_c\rangle\propto l^d$, corresponding to fractal
 dimensions $d=2\nu=1.30$ and $d=2\nu^*=1.26$ 
from the fractal Weyl law of Fig.~\ref{fig3} respectively.
}
\label{fig4}
\end{center}
\end{figure}

The network size $N$ grows with the version number of Linux Kernel
corresponding to its evolution in time.
The dependence of $N_\lambda$ on $N$, shown in Fig.~\ref{fig3},
clearly demonstrates the validity of the fractal Weyl law
with the exponent $\nu \approx  0.63$ for ${\bf G}$
(we find $\nu^* \approx 0.65$ for ${\bf G^*}$). 
We take the values of $\nu$ for $\lambda=0.1$ where the number 
of eigenvalues $N_\lambda$ gives a better statistics. 
Withing statistical errors the value of $\nu$ is not sensitive
to the cutoff value at small $\lambda$. The matrix ${\bf G^*}$
has slightly higher values of $\nu$. These results show that the PCN
of Linux Kernel has a fractal dimension $d=2\nu \approx 1.26$
for ${\bf G}$ and $d=2\nu \approx 1.3$ for ${\bf G^*}$.

\section{Fractal dimension of Linux Kernel network}

To check that the fractal dimension of the PCN indeed has this value
we compute the dimension of the network by another direct method
known as the cluster growing method 
(a description of this method can be find in \cite{fractal}).
In this method the
average mass or number of nodes $\langle M_c \rangle$ 
is computed as a function of
the {\it network distance $l$} counted from an initial
seed node with further averaging over all seed nodes.
In a dimension $d$ the mass $\langle M_c \rangle$
should grow as $\langle M_c \rangle \propto l^d$
that allows to determine the value of $d$ for a given network.

It is important to note that up to now the 
cluster growing method (see e.g. \cite{fractal})
was used only for undirected networks.
Our network is a directed one and, since we did not find
relevant methods for such a case in a literature, we generalized this
method in a simple way using  the network distance $l$
following only outgoing links. The average 
of $\langle M_c(l) \rangle$ is done over all nodes.
Due to global averaging the method gives
the same result for the matrix with inverted link direction
(indeed, the total number of outgoing links is equal to the number
of ingoing links). We will see that the fractal 
dimension obtained by this generalized method
is very different from the case of converted undirected network.

\begin{figure}[h!]
\begin{center}  
\includegraphics[width=.475\textwidth]{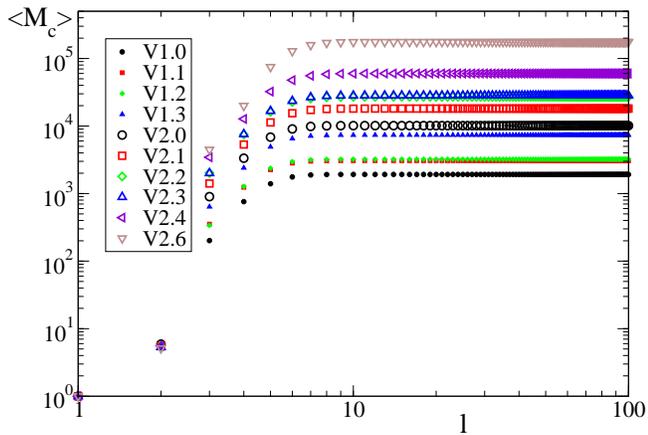}
\caption{(Color online) Cluster growing function for the 
undirected version of the Linux Kernel networks. 
The average mass $\langle M_c\rangle$, with uniformly 
distributed seed, is shown for 
different undirected Kernel versions as a function 
of the \emph{network distance} $l$, in logarithmic scale.
}
\label{fig5}
\end{center}
\end{figure}

\begin{figure}
\begin{center}  
\includegraphics[width=.475\textwidth]{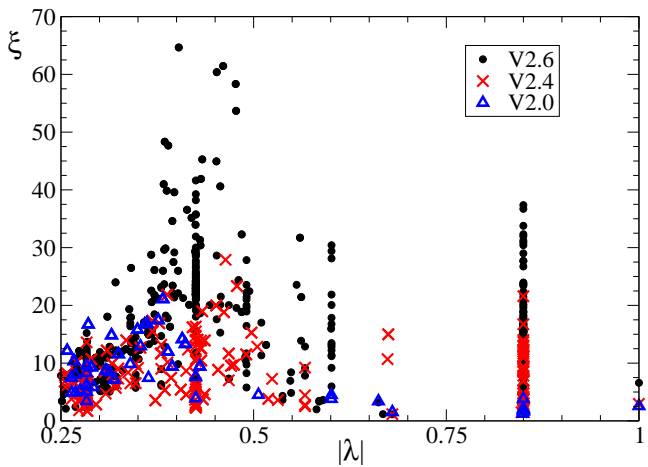}
\caption{(Color online) Localization properties 
of Google matrix eigenstates 
for Linux Kernel versions $2.0.40$ $2.4.37.6$ and $2.6.32$
with $N=14079$, $85756$, $285509$ respectively.
The participation ratio $\xi$ of eigenstates 
is shown as a function of  modulus of 
the corresponding eigenvalues $|\lambda|$;
the values of $<\xi>$, averaged over all given interval of $|\lambda|$, are
$9.5, 8.6, 17.3$ respectively.
}
\label{fig6}
\end{center}
\end{figure}

The dependence of $\langle M_c \rangle$ on $l$,
obtained for the directed network, is shown in Fig.~\ref{fig4}.
Up to a saturation regime related to a finite number of nodes
the growth of $\langle M_c \rangle$ is satisfactory described 
by a power law with a certain fractal dimension $d$.
The straight lines, with the value of $d$ obtained from the 
fractal Weyl law, give in average a satisfactory description of
the growth $\langle M_c(l)\rangle $.
More detailed data, 
with the fitted values of $d$ for the interval $1 \leq l \leq 10$,
are shown in the inset panel of Fig.~\ref{fig4}
for various  network sizes $N$.
For Linux versions up to V2.3 the fractal 
dimension remains in the interval
$1.3 \leq d < 1.45$ with an average $d \approx 1.4$.
However, we have $d \approx 1.6$ for V2.4
and $d \approx 1.9$ for V2.6 versions. 
We attribute the strong deviation for the version V2.6
to the well known fact that significant 
rearrangements in the Linux Kernel have been done
after version V2.4 \cite{linuxva,linuxvb}.
Also the data of Fig.~\ref{fig4} clearly show the saturation of 
$\langle M_c(l)\rangle $ at relatively small values
$\langle M_c(l)\rangle < 200$ compared to the whole size of the
network $N \leq 285509$. This happens due to 
existence of many isolated communities in the network
which become linked only due to regularization of the Google
matrix (\ref{eq1}). This is the reason of strong
degeneracy of the eigenvalue $\lambda=\alpha$
which has $116$ values for the version V2.6.
Also, in contrast to undirected networks, 
in directed networks there are
nodes with only ingoing links that give an important
reduction of the average mass.

Thus in view of the above restrictions we consider that there is
a rather good agreement of the fractal dimension
obtained from the fractal Weyl law with $d \approx 1.3$
and the value obtained with the cluster growing method
which gives an average $d \approx 1.4$.
The fact that $d$ is approximately the same for
all versions up to V2.4  means that the  Linux Kernel
is characterized by a self-similar fractal
growth in time. The closeness of $d$  to unity
signifies that procedure calls are almost linearly ordered 
that corresponds to a good code organization.
Of course, the fractal Weyl law gives the dimension $d$
obtained during time evolution of the network.
This dimension is not necessary the same as 
for a given version of the network of fixed size.
However, one can expect that the growth 
goes in a self-similar way \cite{dorogovtsev,havlinnphys} 
and that static dimension is close to the 
dimension value emerging 
during the time evolution. This  can be viewed 
as a some kind of ergodicity 
conjecture. Our data show that this conjecture
works with a good accuracy up to the Linux Kernel V.2.6.

Here it is important to note that the
properties of the directed network are very different from the
converted undirected network with the same nodes.
The later is obtained from the directed network by
replacing all directed links by undirected ones. 
The results obtained by the clustering growing
method  for this undirected network
are shown in Fig.~\ref{fig5}. We find that it is difficult to
fit the dependence $\langle M_c(l)\rangle$ by a simple power 
law in this case. A very rough estimate gives very
high value $d \sim 5$ being by a factor 4 larger
than the dimension of the directed network.
Also the saturation level of $\langle M_c\rangle$
is now comparable with the whole system size
$N$ while for the directed network
it is smaller than $N$ by three orders of magnitude.

Usually the methods of computation of fractal dimension
have been developed for undirected networks
(see e.g. \cite{fractal,havlinnphys,eplnet,kim}).
Our example shows that the directed networks
may have rather different characteristics compared to their
undirected versions. Thus new methods for computation of 
fractal dimension in directed networks should be 
developed. Our results show that the generalized version
of the method described in \cite{fractal}
work well in the case of directed networks.

\begin{figure}[h!]
\begin{center} 
\includegraphics[width=.483\textwidth]{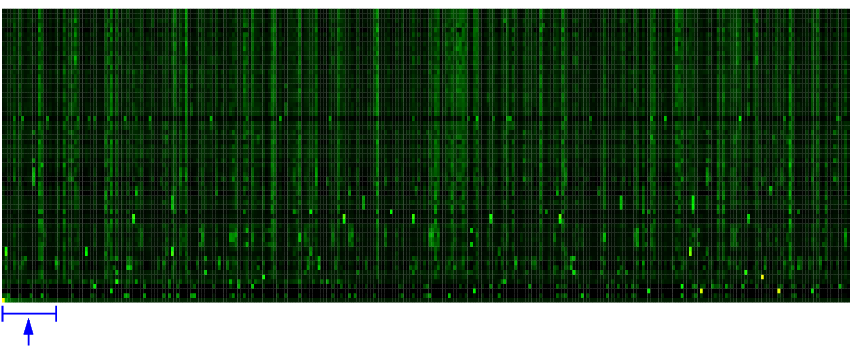}
\includegraphics[width=.483\textwidth]{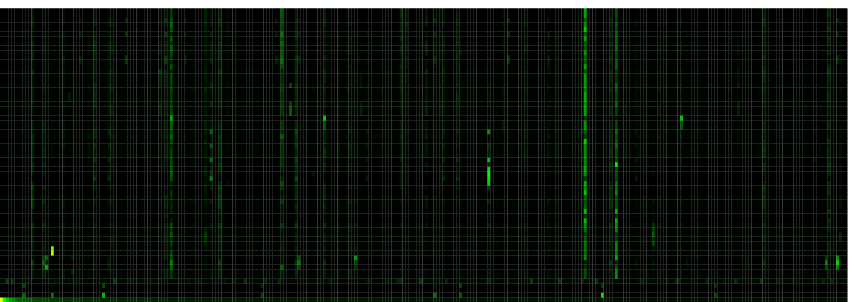}
\caption{(Color online) 
Coarse-grained probability distribution $|\psi_i(j))|^2$ for the
eigenstates of the Google matrix for Linux Kernel version $2.6.32$.
The horizontal lines show the first $64$ eigenvectors ordered 
vertically by $|\lambda_i|$ 
from $\vert\lambda\vert=1$ on bottom 
to $\vert\lambda\vert \approx 0.4$ on top
(only one state is 
shown for degenerate eigenvalues, all states are normalized by 
$\sum_j\vert\psi_i(j)\vert^2=1$, index $j$  corresponds to 
the order given by the  PageRank located at the bottom line).
The top panel shows the coarse-grained complete network 
of 285509 sites with $307$ cells, each containing $930$ sites; 
the bottom panel shows the coarse-grained distribution 
for first $300$ cells, each containing $62$ sites
(bar and arrow on the top panel mark the region of bottom panel).
Probability is proportional to color with
black for zero and yellow/white for maximum.
}
\label{fig7}
\end{center}
\end{figure}

\section{Google matrix eigenstates}

Before we discussed the properties of the spectrum of the Google matrix 
of the PCN. Let us now analyze  the properties of 
its eigenstates $\psi_i(j)$. As it is usual for the disordered systems with 
the Anderson localization  (see e.g. \cite{mirlin}), 
we use the PArticipation Ration (PAR)
$\xi$ to characterize the localization properties of the eigenstates.
It is defined as $\xi_i=(\sum_j|\psi_i(j)|^2)^2/\sum_j|\psi_i(j)|^4$.
Physically, $\xi_i$ gives the number of nodes effectively
occupied by an eigenstate $\psi_i(j)$. The dependence
of $\xi$ on $|\lambda|$ for three versions of the Linux Kernel
is shown in Fig.~\ref{fig6}. There is only a small increase of $\xi$
with size $N$  which is however much smaller
than the increase of $N$ from one version to another.
Also even maximum values of $\xi$ remain by a factor $5 \times 10^3$
smaller than the system  size $N$. Thus we conclude that the eigenstates 
are essentially localized.

The dependence of probability of eigenstates,
with largest values of $|\lambda_i|$, on the PageRank index $j$ 
is shown in Fig.~\ref{fig7} for the version 2.6.32.
The main peaks of probability are scattered over the whole system
size $N$. The appearance of vertical stripes is clearly visible.
This signifies that statistically the positions of peaks are repeated from
one eigenstate to another. It is possible that this effect
is related with certain cycles between procedures
in the Linux Kernel. These cycles, or communities as 
it is used to say for the WWW,
appear even on a far tail of the PageRank.
The origin of such weakly isolated cycles and correlations 
between different relaxation modes requires further detailed studies. 

\section{Discussion}

Here we characterized the directed network by the Google matrix 
and investigated its spectrum and eigenstate properties.
We think that such an approach gives much richer 
charcteristization of  networks. Indeed, 
as it was shown in \cite{georgeot}, the spectrum of the Google matrix
depends in a sensible way on network structure
so that networks with rather similar distribution
of links may have very drastic differences in the spectrum
(e.g. presence or absence of gap in a vicinity of $\lambda=1$).
We think that further studies of the Goggle matrix properties of
complex networks will bring us deeper understanding
of their generic properties.

In conclusion, our results show that the Google matrix of the
PCN of the Linux Kernel is characterized by the fractal Weyl law
with the fractal Weyl exponent $\nu \approx 0.65$. This value corresponds 
to the fractal dimension of the network $d \approx 1.3$.
This dimension is smaller than two and thus the fractal Weyl law becomes
well visible. In contrast to that for networks with the
dimension $d \geq 2$ the fractal Weyl law is
replaced by a  usual dependence $N_\lambda \propto N$
since $\nu =d/2 >1$. It is known that the WWW has the fractal dimension
$d \approx 4.1$ \cite{fractal} that explains why the fractal 
Weyl law does not work for the WWW university networks 
discussed in \cite{georgeot}. On the basis of our result 
we argue that the fractal Weyl law
should appear in all directed networks with the fractal dimension $d < 2$.
It is important to note that the fractal Weyl exponent $\nu$
is not sensitive to the exponent characterizing the decay of the PageRank:
indeed, according to \cite{alik}
the later remains the same for the WWW and the PCN of Linux Kernel
while the values of $\nu$ are different.
We think that the further studies of the fractal Weyl law for directed
networks with $d<2$ will bring us to a deeper understanding of their 
scale-free  properties. Our results also show that the properties 
of directed networks can be rather different from
their undirected versions.

\end{document}